\documentstyle[11pt,aaspp4]{article}
\begin{document}

\newcommand\Min{{\rm min}}
\newcommand\Max{{\rm max}}
\newcommand\nhat{\hat{n}}
\newcommand\vth{v_{\rm th}}
\newcommand\oneover[1]{\frac{1}{#1}}
\newcommand\parens[1]{\left(#1\right)}
\newcommand\pfrac[2]{\parens{\frac{#1}{#2}}}
\newcommand\partials[2]{\frac{\partial{#1}}{\partial{#2}}}
\newcommand\Civ{C{\sc iv}}
\newcommand\Lya{Ly$\alpha$}
\newcommand\Nv{N{\sc v}}
\newcommand\Ciii{$\left.\mbox{C{\sc iii}}\right]$}
\newcommand\Hb{H$\beta$}
\newcommand\Heii{He{\sc ii}}

\title{Reverberation Mapping and the Disk Wind Model of the Broad Line
Region}

\author{J.~Chiang and N.~Murray}
\affil{Canadian Institute for Theoretical Astrophysics, McLennan Labs,
60 St. George Street, Toronto, Ontario, Canada, M5S 1A1}

\begin{abstract}
Using the disk wind model of Murray et al. (1995), we calculate line
profiles and frequency-resolved response functions for broad line
emission from the surface of an accretion disk in an AGN in the
presence of a radiatively driven wind.  We find that the combined
effects of the shears in the wind and in the disk itself produce
anisotropic line emission which solves several well-known problems
connected with disk models of the broad line region.  In particular,
the broadening of resonance lines such as \Civ, \Lya, and \Nv\/ can
be attributed to orbital motion of the disk gas at radii as close as
$\sim 10^{16}$~cm in Seyferts without requiring unrealistically
large emission regions in order to produce single-peaked profiles.
Furthermore, the anisotropy of the line emission results in
frequency-dependent response functions which are no longer red-blue
symmetric so that the time delays inferred for the various red and
blue components of the line agree qualitatively with recent
reverberation mapping observations of NGC~5548.
\end{abstract}


\section{Introduction}

Broad emission lines (BELs) are the distinguishing feature of AGN
spectra; and the variability, absorption, profile shape and strength
of these lines are our most powerful probes of the inner structure of
active galactic nuclei.  The basic photoionization model and the
evidence for variability in the both the lines and continuum fluxes of
AGNs led Blandford \&~McKee (1982) to propose ``reverberation
mapping'' as a means of constraining the geometry and kinematics of
the Broad Line Region (BLR).  The principle underlying reverberation
mapping is that the light curve of a given line, $L(t)$, can be
related to the continuum light curve, $C(t)$, by a causal, linear
transformation.  The frequency-dependent version of this
transformation is
\begin{equation}
L(t,\nu) = \int_0^{+\infty} \psi(\tau,\nu) C(t-\tau)\,d\tau,
\end{equation}
where the transfer function $\psi(\tau,\nu)$ is the emission line
response to a $\delta$-function outburst of the ionizing continuum.  In
principle, the convolution theorem and Fourier inversion can be used
to determine $\psi(\tau,\nu)$ from data.  The actual geometry and
velocity structure of the BLR can then be inferred from some model for
the kinematics of the BLR and the fact that the surfaces of constant
delay are paraboloids aligned along the observer line-of-sight with
the central continuum source at the foci.

The difficulties associated with actually performing such an inversion
have motivated long-term observations of AGN in order to obtain high
quality spectra and light curves.  The first of these extensive
monitoring programs was the combined IUE (International Ultraviolet
Explorer)/ground-based observations of NGC~5548 in 1988-89 (Clavel et
al. 1991).  Several major insights into the BLR resulted from the
analysis of these NGC~5548 observations.  First of all, the size of
the BLR was found to be substantially smaller than the size indicated
by photoionization calculations of cloud models of the BLR (Krolik et
al 1991).  Furthermore, cross-correlation analysis of light curves of
various lines with the continuum has shown that the BLR is stratified:
longer lag times were found for lower ionization lines such as \Hb\/
and \Ciii\/ than for higher ionization lines such as \Nv, \Civ\/ and
\Heii\/ (Krolik et al. 1991).  Also, because of the higher quality
spectral data, the response functions of separate components of a
single line could be studied.  Crenshaw \&~Blackwell (1990) found
evidence for a faster response from the red-shifted half of the \Civ\/
line versus the blue-shifted half.  The complexity of the BLR which
these studies revealed called for even better data.

The recent HST/IUE/ground-based effort to monitor NGC~5548 (Korista et
al. 1995; hereafter K95) has produced the highest quality data yet
obtained for probing the BLR via reverberation mapping techniques.
K95 have confirmed the difference between the red-shifted and
blue-shifted responses of the \Civ\/ line.  They have also provided
better constraints on the ionization stratification for some of the
higher ionization state lines.

The most direct evidence of these effects has been obtained by
examining the cross-correlation of the emission line light curves with
the continuum.  Recently, efforts have been made towards
deconvolutions of the actual line response functions using various
techniques (Horne 1994; Pijpers 1994; Krolik \&~Done 1995).
Underscoring the importance of modeling, ``echo images'' in the
$\tau$-$\nu$ plane for various models of the broad line region
(including Keplerian disks and spherically symmetric models with
radial infall or out-flow) have been produced, anticipating
frequency-resolved response function deconvolutions (Welsh \&~Horne
1991; Perez, Robinson, \&~de~la~Fuente 1992).  Done \&~Krolik (1995)
and Wanders et al. (1995) have made the first attempts at such a
deconvolution for the \Civ\/ line of NGC~5548.  Following K95, they
divided the line into four frequency bands---two ``core'' components
and two ``wing'' components; they derived separate response functions
for each, confirming the results of the cross-correlation results of
K95, and were able to study specific BLR models in greater detail.

Current thinking about the BLR has been dominated for the past decade
by the cloud picture (see Peterson 1994 for a review).  In these
models, dense clouds appropriately distributed in velocity are
photoionized by the central continuum and emit the line photons which
form the broad lines.  Since there is very little agreement about how
such clouds are formed, how they are confined, and how they attain
their velocity distribution, cloud models have the advantage of
considerable flexibility in describing the observations.
Reverberation mapping and photoionization models have placed some
constraints on the nature of the BEL cloud region: the characteristic
radius of the BLR is likely $\sim 3$-$10$ lt-days; the clouds are
almost certainly optically thick, though the absence of Lyman edges
suggests that individual clouds are significantly smaller than the
size of the central continuum emitting region; the smoothness of the
line profiles implies that there are of order $> 10^5$ clouds
contributing to any given line profile.  Furthermore, if the clouds
are distributed spherically or nearly so, then the tendency for
non-zero lags in the line response implies that the cloud emission is
anisotropic, with the emission directed primarily towards the central
source.  This is consistent with the clouds being optically thick, and
Ferland et al. (1992) have performed photoionization calculations to
determine the expected emission profile for such clouds.  Finally, the
faster response times for the red wing of the \Civ\/ line of NGC~5548
versus the blue wing imply that any radial motion which the clouds
possess must be predominantly inward rather than away from the central
source.

It is this last feature which has led some to dismiss the disk model
of the BLR.  Because the velocity field in a disk, which is usually
taken to be Keplerian, is symmetric with respect to red and blue
shifted emission, it is expected that the timing response for both
sides of the line should also symmetric (Done \&~Krolik 1995).
However, this is only necessarily the case if the line emission from
the surface of the disk is {\it isotropic} as is generally assumed
(Welsh \&~Horne 1991; Dumont et al. 1990; Blandford \&~McKee 1982).
In this work, we show that in the disk wind model of Murray et
al. (1995; hereafter M95) the combination of the shears in a Keplerian
disk and a radiatively driven disk wind produces an anisotropic
opacity for the line emission.  This emission anisotropy has several
important consequences: it solves the double-peak problem of disk line
profiles (Mathews 1982) without requiring unrealistically large BLR
radii; it provides a natural explanation for the faster response of
the red-shifted line emission despite the symmetry in the disk
velocity structure; and it establishes a connection with the observed
blueshifted absorption in the \Civ\/ line of NGC~5548 and the disk wind
models of Broad Absorption Line QSOs. Furthermore, this model for the
broad line region allows the mass of the central object to be
constrained, and thus provides additional evidence for the massive
black hole hypothesis for the centers of active galactic nuclei.

The plan for the remainder of this paper is as follows: In \S~2, we
review the disk wind model of M95, discuss its observational
connection with warm absorbers and Seyfert galaxies, derive the
expressions for computing time-averaged line profiles, and compare a
model line profile with data from NGC~5548.  In \S~3, we derive the
frequency-resolved response functions, discuss how the red/blue timing
asymmetry arises, apply our calculations to the NGC~5548 data and
compare our results with the deconvolved response functions of Done
\&~Krolik (1995) and Wanders et al. (1995).  In \S~4, we discuss some
observational consequences of our model, address possible objections
and discuss prospects for more detailed work in the future.  Finally,
we summarize our conclusions in \S~5.

\section{The Disk Wind Model}

Our model for the BEL region of AGNs is based upon the Broad
Absorption Line (BAL) QSO Model of M95.  In BALQSOs, the broad line
spectra are similar to that of ordinary QSOs except for the deep,
broad absorption troughs blueward of line center.  Based on this
similarity, Weymann et al. (1991) have concluded that ordinary QSOs
and BALQSOs belong to the same class of object---only the observer's
line-of-sight to the central source distinguishes them, passing
through absorbing material for the latter objects.  Because some of
the line emission is also absorbed, the cloud picture requires that
two distinct sets of clouds exist in the broad line region.  The broad
emission line clouds exist at the inner radii, closer to the central
continuum, while the broad absorption line clouds surround them, with
appropriate covering factor to allow the absorption line
emission.\footnote{Various descriptions of cloud models for BEL and
BAL regions can be found in Arav et al. (1994); Emmering, Blandford,
\&~Schlossman (1992) and references therein.}

The complexities and difficulties associated with cloud models of
BALQSOs led M95 to consider a different approach.  Noting the
similarity in the BAL profiles to the P-Cygni profiles found in O~star
spectra, they considered the possibility that a radiatively driven
wind provides the observed absorption. In O~stars, it has been well
established that such winds are responsible for the shape of the
resonance lines (Castor, Abbott \& Klein 1975; Pauldrach et al. 1994).
Following the formalism of Castor et al. (1975) for radiatively driven
winds, M95 calculated the dynamics of a wind which is driven from the
surface of an accretion disk.  They found that the velocity and
ionization structure of these winds can account for the width and
shape of the absorption troughs as well as providing a natural
explanation of the inferred covering factor ($\sim 10$\%) of the BAL
region.  The relevance of this model for lower luminosity AGN became
clearer when a connection was established between the discovery in
X-rays of photoionized bulk outflows, the so-called ``warm absorbers''
(Turner et al. 1993; Mathur et al. 1994), in Seyfert galaxies, and the
presence of blueward absorption features in resonance line spectra in
these same objects (Murray \&~Chiang 1995a).  These results suggest
that radiatively driven winds may be a ubiquitous feature of all AGN.
Since broad emission lines are also universal, it is natural that a
connection exists between them as well.

In order to illustrate the applicability of the disk wind model to the
broad emission line region, we consider a simplified description of
the velocity structure of the wind.  The wind consists of nearly
radial streamlines which have footpoints at various radii on the
surface of the disk.  Near the footpoint radius, $r_f$, for a given
streamline, the gas is initially driven upwards by the surface
radiation of the disk.  At a vertical height which is approximately
the scale height of the disk, the radiation pressure from the central
continuum source becomes strong enough to push the gas predominantly
radially outwards.  M95 have integrated the equations of motion using
the radiation force law determined from calculations using the
photoionization code CLOUDY (Ferland 1993) and the line list of
Verner, Barthel, \& Tytler (1994).  They found that the radial
component of the gas velocity along a streamline can be described
approximately by
\begin{equation}
v_r(r) = v_\infty \left(1 - \frac{r_f}{r}\right)^\gamma, \label{radial_vlaw}
\end{equation}
where the terminal velocity of the streamline, $v_\infty$, is within
factors of order unity of the local escape velocity, $v_{\rm esc} =
\sqrt{2GM/r_f}$, and the exponent $\gamma \simeq 1.0$-$1.3$.  It
should be noted that this exponent differs from the result for O~stars
($\gamma\simeq 0.5$, Castor et al. 1975) because of the differing
force multiplier law, as derived from the photoionization
calculations, and the importance of the centrifugal support of the
rotating Keplerian disk.  In the following discussion, we will use
$\gamma = 1$.

M95 have performed dynamical and photoionization calculations which
show that the inner edge of the wind should be located at $\sim 3
\times 10^{15}$cm for an accretion disk around a $10^7\,M_\odot$ black
hole.  It is important to note that at these radii, the shear in the
disk itself is sufficient to make the Sobolev approximation an
appropriate description of the resonance line transfer in the disk (cf
Castor 1970).  Murray \&~Chiang (1995b) discuss the validity of this
approximation in greater detail.  For our purposes, it is sufficient
to note that the Sobolev length scale associated with the shear due to
the differential rotation of the disk is
\begin{equation}
l_S \sim \frac{\vth}{v_\phi/r} \sim 10^{10}\,\mbox{cm},
\end{equation}
where $\vth \sim 10^6$~cm~s$^{-1}$ is the thermal velocity of the disk
gas, and $v_\phi = \sqrt{GM/r}$ is the azimuthal disk velocity.  This
length is much smaller than the length scale of the Shakura-Sunyaev
(1973) thin-disk solution of $\sim 10^{13}$~cm for a $10^7\,M_\odot$
black hole; this implies that, as within the wind itself, the Sobolev
approximation also applies to the disk.

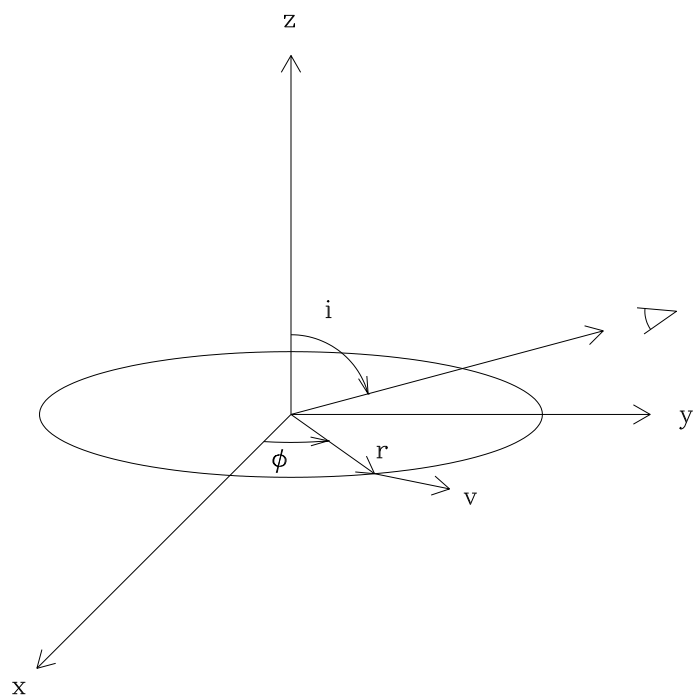
\begin{figure}
\caption{Geometry of the disk broad line region.  The angle, $i$, is
the disk inclination relative to the observer.  The quantities
$(r,\phi)$ label locations on the disk.}
\label{DiskSchematic}
\end{figure}

We expect that the largest contribution to the line emission will be
from regions of high density.  Therefore, we will assume for now
that the line emission originates primarily near the wind footpoints
at the surface of the accretion disk where the densities are high
($n_H \sim 10^{9}$cm$^{-3}$), but the continuum optical depths, which
are largely due to electron scattering, are still no more than order
unity.  At these locations, the velocity of the gas is composed of a
radial component due to the wind and an azimuthal component due to the
disk motion:
\begin{equation}
\vec{v} = v_r \hat{r} + v_\phi \hat{\phi} \label{disk_velocity}.
\end{equation}
For an observer located in the $y$-$z$ plane with a line-of-sight
angle $i$ with respect to the disk symmetry axis (see
Figure~\ref{DiskSchematic}), the projected velocity shift at a
location $(r,\phi)$ on the disk is
\begin{equation}
v_{\rm proj} = \sin i(v_r\sin\phi + v_\phi\cos\phi).
\label{vproj}
\end{equation}
Note that we have neglected the finite thickness of the disk ($\sim
10^{13}$~cm) which is much smaller than the inferred radius of the BLR
($\sim 10^{16}$~cm).

Because of the azimuthal symmetry of the problem, we model the line
source function, $S(r)$, to be a function only of the disk radius.
The contribution to the line profile at a given frequency is
\begin{equation}
L_\nu = \cos i \int_{r_\Min}^{r_\Max} r dr\,\int_0^{2\pi} d\phi\,
    k(r) S(r) \frac{1 - e^{-\tau}}{\tau} \delta(\nu - \tilde{\nu}(\phi,r)),
\label{LnuIntegral}
\end{equation}
where $r_\Min$ and $r_\Max$ are the radial boundaries of the BLR,
$\tilde{\nu}$ is the Doppler-shifted frequency of the line at the disk
element which has velocity $v_{\rm proj}$:
\begin{equation}
\tilde{\nu} = \nu_0\left(1 + \frac{v_{\rm proj}}{c}\right)
\end{equation}
and $\nu_0$ is the frequency of the line.  The optical depth $\tau$
is angle dependent and is {\it locally} determined in the Sobolev
approximation, and $k(r)$ is the integrated line opacity (Rybicki
\&~Hummer 1983).
In the notation of Hamann, Korista \&~Morris (1993)
the quantity
\begin{equation}
\beta(r,\phi,i) = \frac{1-e^{-\tau}}{\tau}
\end{equation}
is known as the {\it directional} escape probability.  Transforming the
$\delta$-function and evaluating the $\phi$-integral, we obtain for the
integrand of Equation~\ref{LnuIntegral}
\begin{eqnarray}
\frac{dL_\nu}{dr}
     &= &r\,k(r) S(r) \cos i\left|\frac{d\tilde{\nu}}{d\phi}\right|^{-1}
         \beta(r,\phi,i)\\
     &= &\frac{c}{\nu_0}\frac{r\,k(r) S(r)}
         {\tan i\left|v_r\cos\phi - v_\phi\sin\phi\right|}
         \beta(r,\phi,i)
\end{eqnarray}
where $\phi$ satisfies $\tilde{\nu}(r,\phi) = \nu$ for a given radius
$r$.

The line optical depth in the Sobolev approximation is
\begin{equation}
\tau = \frac{\kappa \rho \vth}{|\nhat\cdot\Lambda\cdot\nhat|},
\label{OpticalDepth}
\end{equation}
where $\kappa$ is the line absorption coefficient, $\rho$ is the
density, $\nhat$ is the line-of-sight vector to the observer and
$\Lambda$ is the symmetric strain tensor (Castor 1970; Rybicki
\&~Hummer 1978).  For the \Civ\/ line, the line absorption coefficient
is
\begin{eqnarray}
\kappa \simeq 2.61 \times 10^7 \eta_{{\rm C}^{+3}} \label{kappa}
\end{eqnarray}
assuming a cosmic abundance of carbon and where $\eta_{{\rm C}^{+3}}$
is the fractional abundance of the C$^{+3}$ ion.  We expect
$\eta_{{\rm C}^{+3}}$ to be of order unity throughout the \Civ\/ BELR.
For the velocity structure given by Equations~\ref{radial_vlaw}
\&~\ref{disk_velocity}, we have
\begin{eqnarray}
\nhat\cdot\Lambda\cdot\nhat
     &= &\sin^2i\left[\partials{v_r}{r}\sin^2\phi
        + \parens{\partials{v_\phi}{r} - \frac{v_\phi}{r}}\sin\phi\cos\phi
        + \frac{v_r}{r}\cos^2\phi\right] \\
     &= & \sin^2i\left[v_\infty\frac{r_f}{r^2}\sin^2\phi
        - \frac{3}{2}\frac{v_\phi}{r}\sin\phi\cos\phi
        + \frac{v_r}{r}\cos^2\phi\right] \label {Shear}.
\end{eqnarray}
Evaluated at the high density regions near the footpoints of the wind
streamlines, this factor is of order $\sim v_\phi/r_f$.  Combining
Equations~\ref{OpticalDepth}, \ref{kappa}, and \ref{Shear}, the
optical depth in the line is of order
\begin{equation}
\tau \sim 10^5 \frac{n_H}{10^9 {\rm cm}^{-3}};
\end{equation}
so we can use
\begin{equation}
\beta \simeq \frac{1}{\tau}.
\end{equation}
The final expression for the luminosity per unit radius is
\begin{eqnarray}
\frac{dL_\nu}{dr} &= & r\,S(r)
       \sin i\cos i\left|\frac{\displaystyle
           \partials{v_r}{r}\sin^2\phi
           - \frac{3}{2}\frac{v_\phi}{r}\sin\phi\cos\phi
           + \frac{v_r}{r}\cos^2\phi}{v_r\cos\phi - v_\phi\sin\phi}\right|
         \label{dLnudr} \\
   &\simeq &S(r)\sin i\cos i\left|
                  \parens{\frac{v_\infty}{v_\phi}\sin\phi
                  - \frac{3}{2}\cos\phi}\right| \label{simdLnudr},
\end{eqnarray}
where the final form is obtained by evaluating Equation~\ref{dLnudr}
near the wind streamline footpoints, $r \simeq r_f$, and neglecting
terms which are proportional to $v_r/v_\phi \la 10^{-2}$.  We note
that the remaining $r$-dependence is contained in the source function
$S(r)$ given that $v_\infty$ and $v_\phi$ both scale as $\sim
r^{-1/2}$.

It is not entirely clear how the central continuum source illuminates
the outer regions of the disk where the broad emission lines are
formed.  The standard picture of AGNs is that the UV and X-ray
continuum emission is emitted from the hot inner regions of the
accretion disk.  However, the Shakura-Sunyaev thin disk solution,
which is thought to apply to these inner regions, gives a disk height
which is essentially constant from the hot inner regions out to the
disk BLR.  Therefore, in the standard thin-disk solution, there would
not be any lines-of-sight from the continuum source to the BLR.

However, in disk models of the BLR, it is conventional to assume that
either that a continuum source exists a moderate distance above the
inner regions of the disk (Matt, Fabian \& Ross 1993) or that there is
a hot, spherical region at the center of the disk (Rokaki \& Magnan
1994).  Murray \& Chiang (1995b) propose a physical explanation for
the latter case wherein resonance line pressure makes an additional
contribution to the opacity at the inner regions of the disk causing
it to ``puff up.''  In any case, for the purposes of the reverberation
mapping study, we will assume that there is a central source of
ionizing continuum which is situated not far above the disk plane ($z
\sim 10^{13}(M/10^7 M_\odot)$~cm), and which can illuminate the disk
BLR.

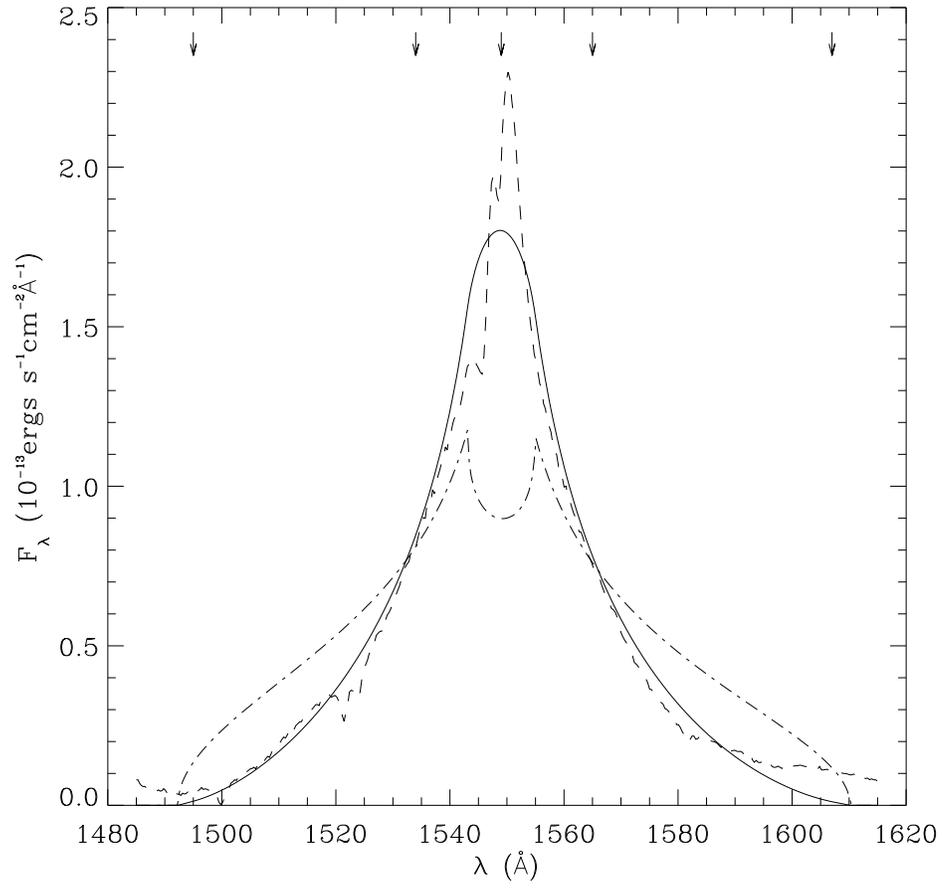
\begin{figure}
\caption{Data and model line profiles for the \Civ\/ line of NGC~5548.
The solid line is the model calculation including the effects of the
anisotropic emission.  The dashed line is the data from the
1994 HST observations described by Korista et al. 1995.  Also shown is
the double-peaked profile (dot-dashed) of a model line calculation
which assumes isotropic emission.  The arrows indicate
the boundaries of the various wing and core components.}
\label{LineProfiles}
\end{figure}

In order to reproduce a reasonable line profile shape (cf
Figure~\ref{LineProfiles}), we have found that the radial dependence
in Equation~\ref{simdLnudr} can be adequately modeled by
\begin{equation}
S(r)\propto r^{-\alpha}, \label{flux_scaling}
\end{equation}
with a value of $\alpha \simeq 1$.  The inner radial boundary of the
BLR, $r_\Min$, is constrained by the reverberation mapping results and
photoionization calculations.  The outer radial boundary, $r_\Max$, is
determined by the location of the Str\"omgren radius for the ion in
question.  Preliminary photoionization results give $r_\Max \sim
10^{2}r_\Min$ for the \Civ\/ line in NGC~5548.  Using this radial
dependence and these boundaries, we have integrated
Equation~\ref{LnuIntegral} and find the line profile (solid line) for
\Civ\/ shown in Figure~\ref{LineProfiles}.  For comparison, the dashed
line is the mean GEX-extracted \Civ\/ line profile of NGC~5548 from
the recent HST observations (K95) and the dot-dashed line results from
the same model calculation except that isotropic emission is assumed.
The effect of the anisotropic optical depth is clear: the disk wind
shears result in single-peaked profiles, and the shape of the line,
except for the blue-ward absorption and a possible contribution from a
narrow-line component, are well-matched.

We have used a disk inclination angle $i = 75^\circ$ for
this calculation.  This choice is motivated by the presence of the
warm absorber and the expectation that it is due to a highly ionized
component of the disk wind which lies in the observer line-of-sight
(Murray \&~Chiang 1995a).  We note that the value of $i$ and the
index of the flux radial dependence given in
Equation~\ref{flux_scaling} are correlated: the \Civ\/ line will be
better fit by values of $\alpha \ga 1$ for smaller values of $i$.
It should also be noted that these results only apply to the {\it
shape} of the line from our model.  Accordingly, the model lines shown
in Figure~\ref{LineProfiles} are normalized so that their equivalent
widths match that of the observed line.  More complete photoionization
calculations would have to be performed using the detailed disk wind
velocity and ionization structure and the actual ionizing continuum of
NGC~5548 for a full comparison with the data to be made.  The
preceding arguments are presented to show that the scalings we have
inferred are reasonable and that there is physical motivation for the
disk wind model.

\section{Disk Wind Response Function}

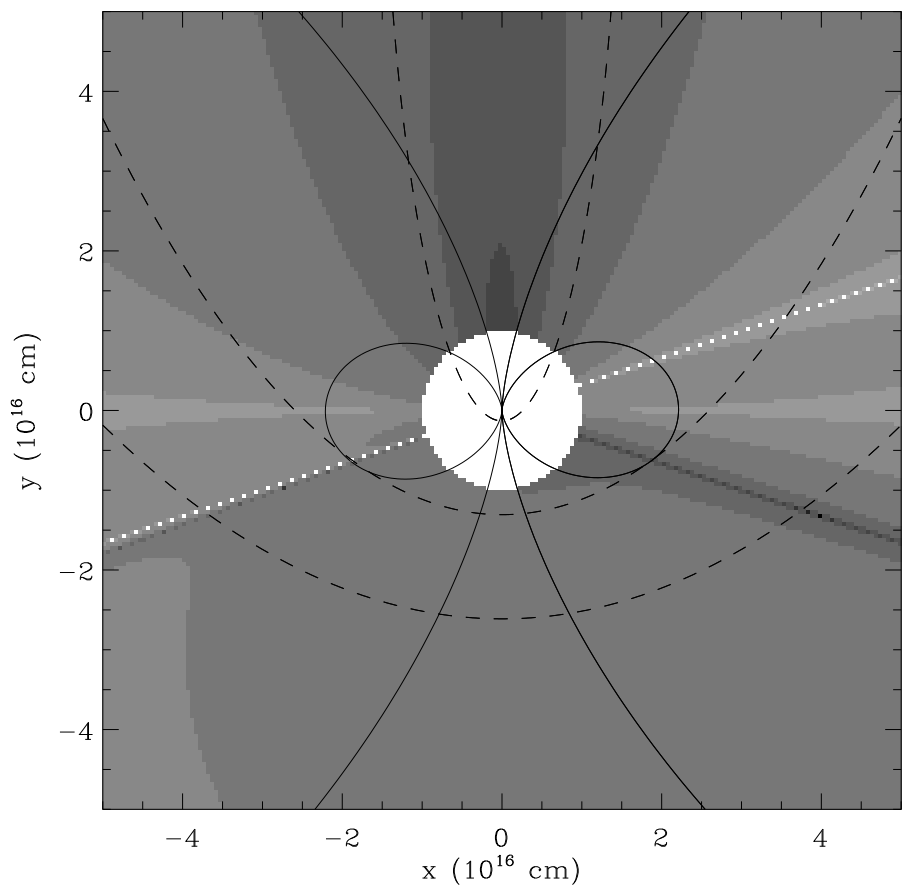
\begin{figure}
\caption{Disk map of the line transfer function $\psi$.  The solid
lines are the lines of constant frequency indicating the boundaries of
the blue and red wing components. They are at 10,840~km~s$^{-1}$
(right side, inner curve), 3,000~km~s$^{-1}$ (right side, outer curve
which extends beyond the plot boundaries), $-3,000$~km~s$^{-1}$ (left
side, outer curve) and $-10,840$~km~s$^{-1}$ (left side, inner curve).
The dashed lines are the lines of constant time delay at $\tau =$0.5,
5 and 10~days.}
\label{DiskMap}
\end{figure}

So far, the line profiles we have presented are for a steady continuum
source.  We now derive the transfer function in order to consider
variability. We have
\begin{eqnarray}
\psi(\tau,\nu) &= &\int_{\rm disk}dA\, k(r)S(r) \beta(r,\phi,i)
    \delta(\tau - \tilde{\tau}(r,\phi))
    \delta(\nu - \tilde{\nu}(r,\phi)) \nonumber\\
&= &\int rdr\int d\phi\,k(r)S(r) \beta(r,\phi,i)
    \delta(\tau - \tilde{\tau}) \delta(\nu - \tilde{\nu}).
\end{eqnarray}
For each point on the disk, $(r,\phi)$, there is associated a
time-delay and frequency, $(\tau,\nu)$.  Transforming the integral
over the disk area to one over $\tau$ and $\nu$, we obtain
\begin{equation}
\psi(\tau,\nu) = \left.\frac{r\,k(r)S(r) \beta(r,\phi,i)}
{\displaystyle\left|
\partials{\tau}{r}\partials{\nu}{\phi}
- \partials{\tau}{\phi}\partials{\nu}{r}\right|}
\right|_{\stackrel{\scriptstyle r = \tilde{r}(\tau,\nu)}
{\phi = \tilde{\phi}(\tau,\nu)}} \label{response_func}.
\end{equation}
The denominator is the Jacobian of the transformation, and the entire
expression is evaluated at the corresponding locations
$(\tilde{r},\tilde{\phi})$ on the disk which yield the desired
time-delay and frequency.  Figure~\ref{DiskMap} is an image depicting
the value of the transfer function as a function of location on the
disk (cf Figure~\ref{DiskSchematic}).  Also plotted are the lines of
constant time delay (dashed lines) and the lines of constant frequency
(solid lines) to illustrate the mapping from $(r,\phi)$ to
$(\tau,\nu)$.  The lines of constant time delay are given by the
intersection of the constant time delay paraboloids and the disk
surface.  Near the disk surface, the radial component of the velocity,
$v_r \sim 10^7$~cm~s$^{-1}$, is small compared with the azimuthal
velocity $v_\phi \sim 10^8$-$10^9$~cm~s$^{-1}$ throughout the BLR.
Therefore the projected velocity (cf Equation~\ref{vproj}) is
approximately
\begin{equation}
v_{\rm proj} \simeq \sqrt{\frac{GM}{r}}\sin i\cos\phi,
\end{equation}
so that for a given frequency shift $\Delta\nu$, we have
\begin{equation}
r \simeq \frac{GM\sin^2 i}{v_{\rm proj}^2}\cos^2\phi.
\end{equation}
{}From this expression, we see that the loci of constant absolute shift
$|\Delta\nu|$ look like lobes which are nearly red-blue symmetric.

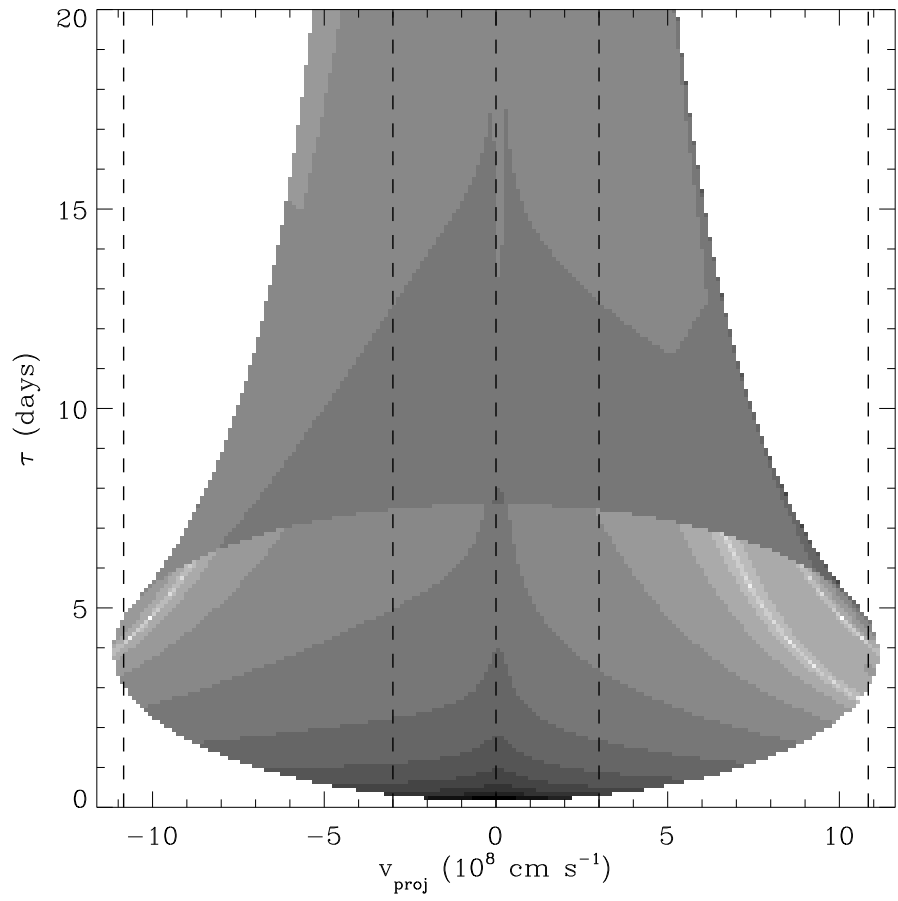
\begin{figure}
\caption{Echo image of the transfer function in $(v_{\rm proj},
\tau)$-space.  The dashed lines are the boundaries of the various line
components.}
\label{EchoImage}
\end{figure}

The red-blue asymmetry of the disk response function can be attributed
to the radiative transfer effects due to the radial and azimuthal
shear of the disk wind.  From Equation~\ref{simdLnudr}, we see that
the $\phi$-dependence of the disk contribution to the line luminosity
is
\begin{equation}
\frac{dL_\nu}{dr} \propto \left|\frac{v_\infty}{v_\phi}\sin\phi
                 - \frac{3}{2}\cos\phi\right|.
\label{phi_dependence}
\end{equation}
Because $v_\infty \sim v_\phi$ and both are intrinsically positive
quantities, the transfer function attains small values near the median
angles ($\sim 45,\,270^\circ$) in the upper right and lower left
quadrants of the disk as shown in Figure~\ref{DiskMap}.  These smaller
values of the emission seen by the observer are depicted by the
lighter pixels in the image.  Figure~\ref{EchoImage} is an echo-image
of the response function in $(\tau,\nu)$-space; note the similarities
and differences with respect to the echo-image maps for disk BLR
models with isotropic emission given in Welsh \&~Horne (1991) and
Perez et al. (1992).

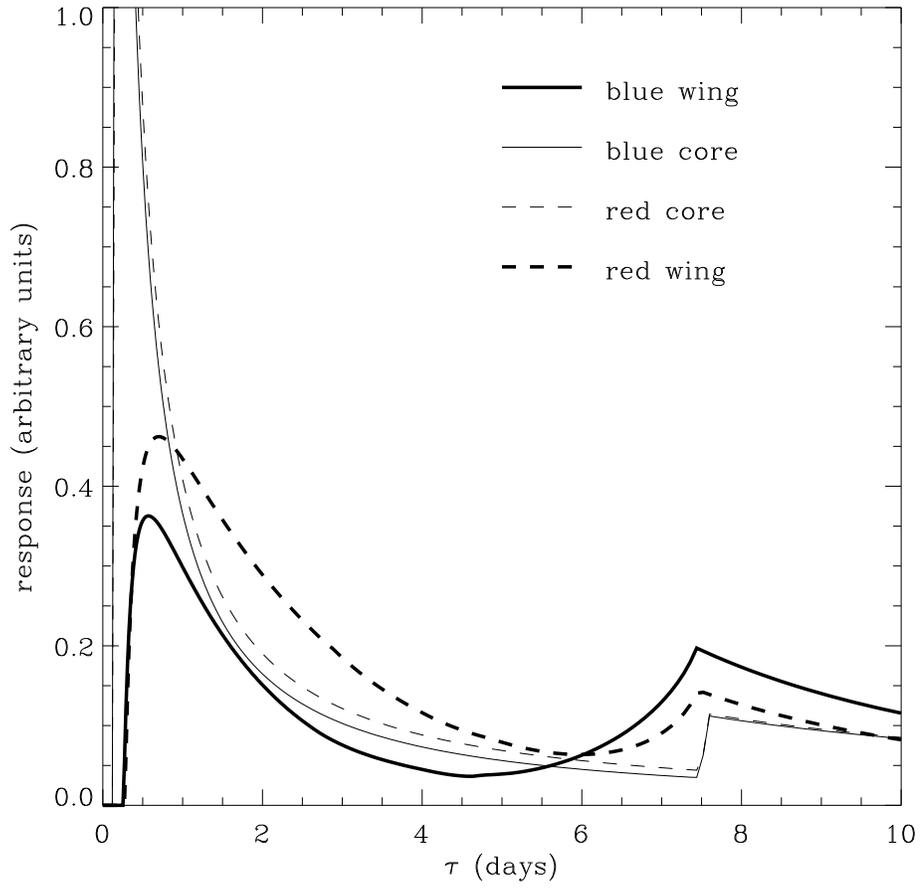
\begin{figure}
\caption{Model response functions for the various components of the
\Civ\/ line.}
\label{ResponseFunctions}
\end{figure}

Following K95, we consider the timing response for the blue/red wing
and core components of the \Civ\/ line.  The core components are
defined to extend from Doppler shifts of 0 to $\pm 3000$~km/s and the
wing components from $\pm 3000$ to $10840$~km/s.  The small arrows in
Figure~\ref{LineProfiles} indicate these boundaries at the
corresponding wavelengths.  We have integrated
Equation~\ref{response_func} over frequency for these four bands and
found the response functions shown in Figure~\ref{ResponseFunctions}.
As with the response functions found by the deconvolutions (Done
\&~Krolik 1995; Wanders et al. 1995), the core components are nearly
identical and are very similar to the blue wing component except at
very short time scales, $\la 1$~day.  Most significantly, the red wing
has significantly stronger response at shorter time scales ($\sim
1$-4~days) than the other three components.  In addition, the
secondary peaks found by both Wanders et al. (1995) and Done \&~Krolik
(1995) at delays of 8-15~days are also a natural consequence of the
disk model.

The faster response of the red wing can be understood by considering
Figure~\ref{DiskMap}.  The blue and red wing frequency ranges are
bounded by the solid lines, and the 0.5, 5 and 10~day time delay lines
are indicated by the dashed lines.  Within the earlier region, between
0.5 and 5~days, the red wing response is stronger than the response of
the blue wing.  This occurs because the locations of weaker emission
due to the near cancellation of two $\phi$-dependent terms in
Equation~\ref{phi_dependence} pass through the blue wing region, but
not through the corresponding red wing region.  At later times, from 5
to 10 days, the situation is reversed---the line of near cancellation
of the $\phi$-dependent terms now passes predominantly through the red
wing region and does not suppress as much emission in the
corresponding blue region.  In effect, the response in the blue wing
``catches-up'' to and surpasses that of the red wing---this is evident
in the second peak of the blue wing response function at $\sim
8$~days.  In Figure~\ref{EchoImage}, the difference in the red and
blue response functions can also be clearly seen.

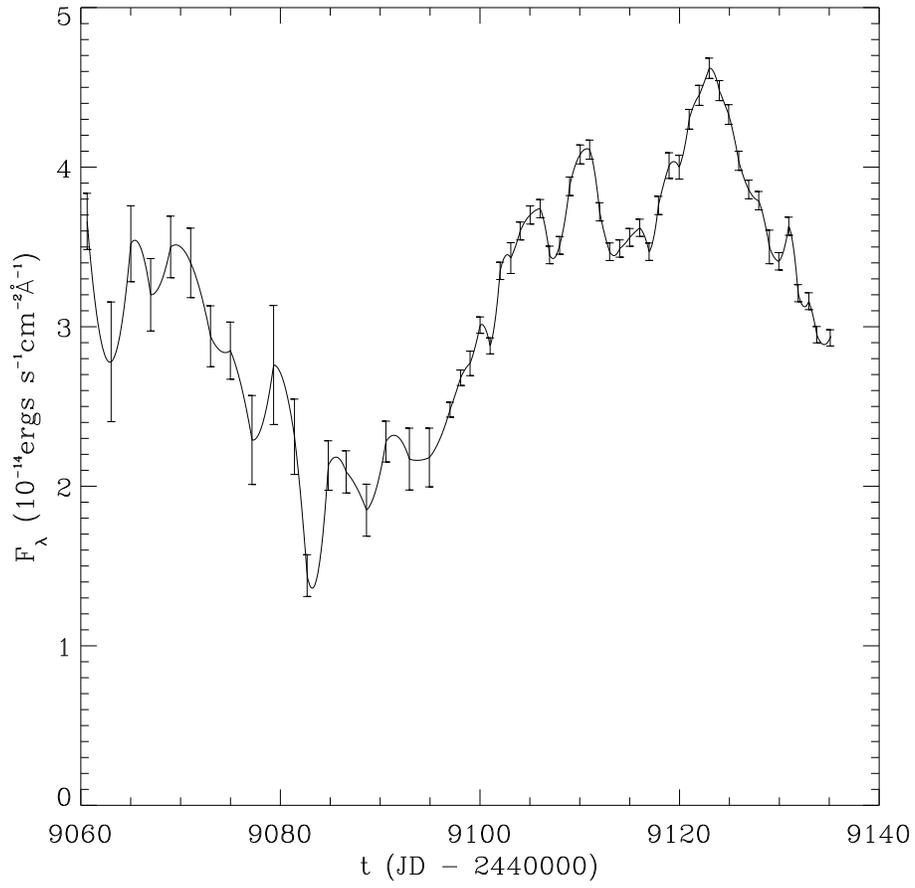
\begin{figure}
\caption{The continuum at 1350\AA\/ of NGC~5548.  The plotted points are
the data as measured by IUE and HST (Korista et al. 1995), and the
solid curve is the interpolated function we have used in our
cross-correlation analysis.}
\label{ContinuumLightCurve}
\end{figure}

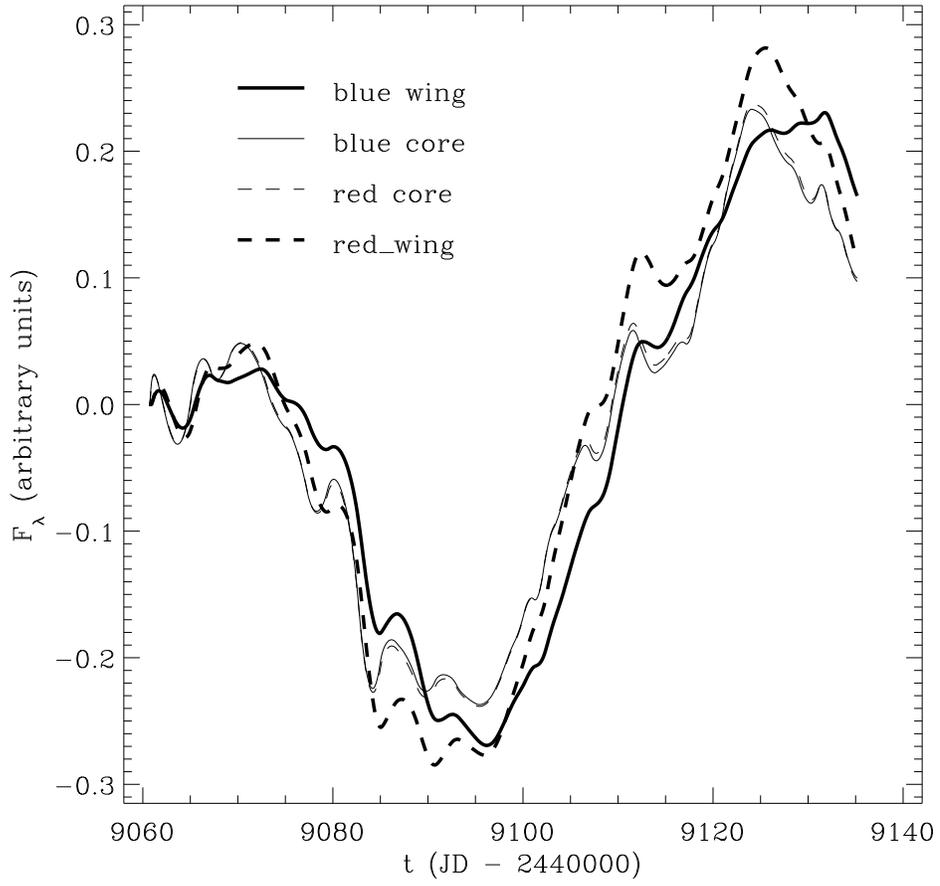
\begin{figure}
\caption{Model light curves of the various components of the \Civ\/ line
of NGC~5548.}
\label{WingCoreLightCurves}
\end{figure}

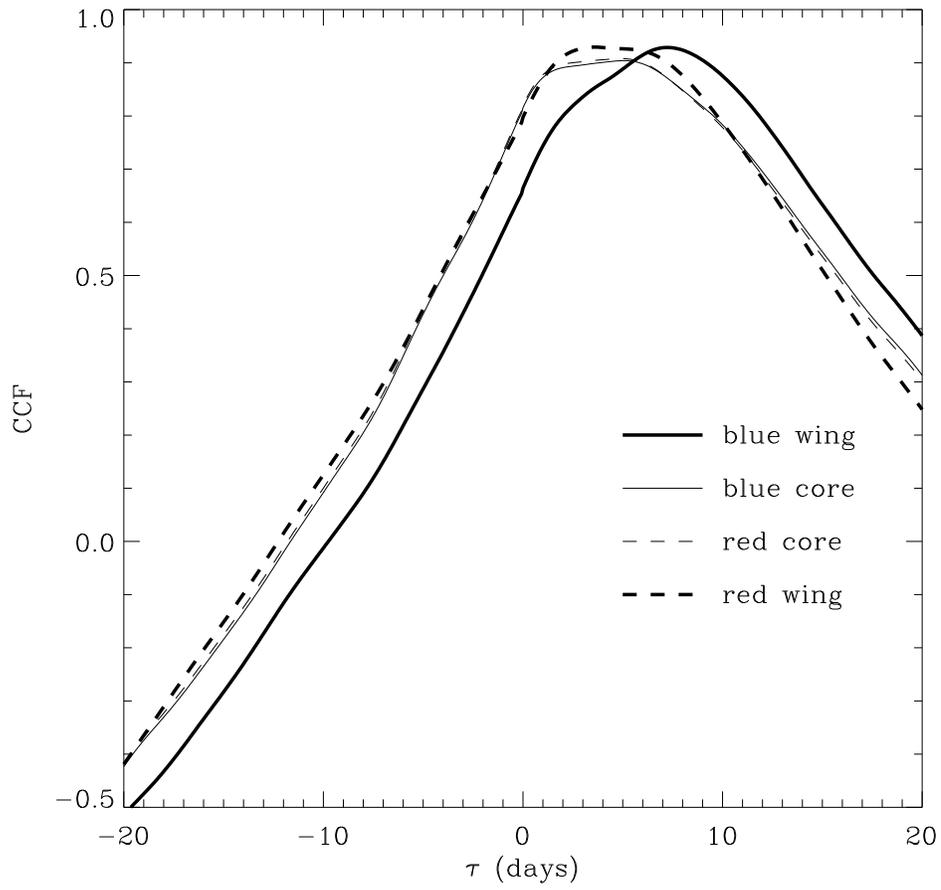
\begin{figure}
\caption{Cross-correlation functions for the four components of the
\Civ\/ line with the 1350~\AA\/ continuum.}
\label{CCFs}
\end{figure}

In order to underscore the applicability of this model to actual data,
we have used the combined IUE/HST spectra of K95 to create light
curves for the four line components and cross-correlation functions of
these model light curves with the continuum.
Figure~\ref{ContinuumLightCurve} shows the 1350~\AA\/ continuum (mean
subtracted) as measured by IUE (SWP) and HST over an $\sim 80$~day
period.  The solid line is the interpolated continuum we have used for
the convolutions, and Figure~\ref{WingCoreLightCurves} shows the
resulting light curves for each of the four line components. The
earlier response of the red wing relative to the blue wing is clearly
evident.  Figure~\ref{CCFs} shows the cross-correlation functions for
these light curves with the interpolated 1350~\AA\/ continuum
(cf Figure~15 in K95). Table~1 shows lists the peak
and centroid values of the inferred lags for each component.

\section{Discussion}

\subsection{Response Function Structure and Line Profile Variations}
The detailed structure of the frequency-resolved response functions
may also be useful in explaining asymmetries in the broad line
profiles as well as long term variability in the line structure.  The
larger second peak in the blue-wing response at $\sim 8$~days may
either ``alias'' or ``anti-alias'' structure present in the continuum
light curve into the mean line profile for a typical observation.  The
result would be line profiles which would skewed toward the either the
blue or the red depending on the phase and strength of the continuum
light curve variations relative to the time window(s) of the
observation.  On the other hand, Perry, van Groningen, \& Wanders
(1994) have shown that the variations seen in the shape of the \Civ\/
line of NGC~4151 are preceded by similar continuum light curves.
They argue that the changes in the line shape may instead be due to
structural changes in the BLR, an extended or anisotropic continuum
source, or a significantly larger BLR than that deduced from
cross-correlation analysis.  In any case, the structure of the
response functions will still be reflected in the shape of the
emission line in the way we have described.  If these other effects
are not significant, then examining the line profile structure and the
preceding continuum variations may be an additional means of
discriminating between different kinematic models of the BLR.

\subsection{Constraints on the Mass of the Central Object}
The radial extent of the BLR is determined by the reverberation
mapping results, and the velocity field is determined by the width of
the resonance lines and the photoionization results for the ionization
structure of the disk wind.  Thus, the observations essentially fix
the quantities $r$ and $v_\phi$.  The assumption of Keplerian motion
relates these two quantities by
\begin{equation}
v_\phi = \sqrt{\frac{GM}{r}},
\end{equation}
so that the mass of the central object is actually constrained by
these observations.  Now, since $M \propto r$ for a fixed line width
(i.e. $v_\phi \sim$ constant) and since the lags are related to radii
on the disk linearly, $\tau \propto r$, we have
\begin{equation}
M \propto \tau.
\end{equation}
The lags which we infer from the cross-correlation functions
(Figure~\ref{CCFs} \&~Table~1) agree with the values given by K95 to
within 20\%, and we have determined the mass of the central object of
NGC~5548 to be $\simeq 10^8\,M_\odot$.  Furthermore, for objects with
line widths similar to NGC~5548, the observed lags for those lines
give a direct measurement of the mass of the central object if this
model for the BLR is correct.

\subsection{Broad Iron K$\alpha$ Lines}
One of the difficulties of this model of the Broad Line Region is that
observations of broad, redshifted iron K$\alpha$ lines in Seyfert 1s
seem to indicate that the disks are being viewed predominantly
face-on with inclinations of $i \sim 20$-30$^\circ$.  In these
objects, the iron emission is believed to originate near the inner
edge of the accretion disk at radii $\sim 3 R_s$, where $R_s$ is the
Schwarzschild radius of the central black hole (Matt et al.~1993).
Mushotzky et al. (1995) report that {\it Ginga} observations of
NGC~5548 have found that the equivalent width of the Fe~K$\alpha$ line
is $\sim 150$~eV and that this is best fit by disk inclinations of
$i \sim 15$-$38^\circ$.  On the other hand, as Done \&~Krolik
(1995) and Wanders et al. (1995) have pointed out, in order for the
response functions of a disk with inner radius $\sim 10^{16}$~cm to
peak near zero delay, the disk must be highly inclined.

We do not believe that these observations of the iron K$\alpha$ lines
and our disk-model for the UV BLR are irreconcilable.  First of all,
the timing resolution of the NGC~5548 observations is no better than
1~day, and we can still produce response functions with the peak
response at time delays $< 1$~day for disk inclinations as small as $i
\sim 60^\circ$.  Secondly, Mushotzky et al. (1995) state that only
three of the five parameters---inner and outer radii of the emitting
region, initial line energy, disk inclination, and line flux---which
are necessary to describe the iron line are actually constrained by
the NGC~5548 data.  Thirdly, Tanaka et al. (1995) and Matt et
al. (1993) have indicated that the emitting material in the Seyfert~1s
may be in a higher ionization state than they have considered and that
this can increase the relative yield of Fe~K$\alpha$ photons, thus
relaxing the constraint on larger inclination angles.  Finally, it
should be noted that the constraints which are also based on the {\it
shape} of the Fe~K$\alpha$ arrived at by Tanaka et al. (1995) for the
Seyfert galaxy MCG-6-30-15 are based upon radiative transfer
calculations which, to our knowledge, ignore the effect of shears in
the disk.  We have seen that these effects can have a significant
impact on the shape of the emission line (cf
Figure~\ref{LineProfiles}) and may also affect the amount of line flux
which escapes.

\subsection{Future Work}
We have not presented here a detailed model of the line emission from
NGC~5548.  The photoionization calculations and our simplified model
of the disk wind structure are too crude to be expected to match the
observations in detail.  Hence, we have not attempted to fit the
NGC~5548 light curves of the \Civ\/ components.  In order for that to
be accomplished, photoionization calculations using a more detailed
physical model of the disk wind structure must be performed.  This
will enable us to determine better the radial dependence for the
emission from each ion, and also probe for non-linear response of the
line emission to the continuum.  We would also like to use more
realistic velocities and shears in Equation~\ref{dLnudr}, and improve
the radiative transfer to take into account possible non-local effects
from multiple resonant Sobolev surfaces (Rybicki \&~Hummer 1978).

\section{Conclusions}
Despite the incompleteness of this model, we have offered a plausible
qualitative explanation of the timing response seen in NGC~5548 using
a physically motivated model for the BLR.  In particular, the response
functions we obtain possess all the features found in the response
functions obtained by the deconvolution procedures of Wanders et
al. (1995) and Done \&~Krolik (1995): they peak near zero time delay;
the red wing response is stronger than the response of the other
components at time delays of $\sim 1$-4~days; and the secondary peak
at 8-14~days is also reproduced.  Furthermore, the line profile which
we obtain is single-peaked in agreement with the profiles seen for most
broad lines in AGN spectra, and we have shown that by using the width
of the line profile and the timing response of the red and blue
components that we can estimate the mass of the central black
hole---for NGC~5548 we find the mass to be $\sim 10^8\,M_\odot$.

The most significant finding of this work is that the radiative
transfer effects which reproduce the observed line profile and timing
response {\it require} that a wind with large radial shears be
present.  Therefore, if it is accepted that the broad lines are
produced from the surface of a disk, then something which acts like a
radiatively driven wind must exist.

\section{Acknowledgements}
We would like to thank Kirk Korista for generously supplying us with
HST spectra of NGC~5548, and Julian Krolik for providing us with a
manuscript of his work in advance of publication.  This work was
supported by NSERC of Canada and by the Connaught Fund of the
University of Toronto.

\begin{table}[p]
\small\centering
\caption{Cross-Correlation Results}
\bigskip
\begin{tabular}{lcccc}
\hline\hline \Civ\/ & \multicolumn{2}{c}{Disk-Wind Model} &
\multicolumn{2}{c}{HST Observations (K95)} \\ component & $\Delta
t_{\rm peak}$ (days) & $\Delta t_{\rm centroid}$ & $\Delta t_{\rm
peak}$ & $\Delta t_{\rm centroid}$\\ \hline
%Blue wing & 6.9 & 5.1 & 7.5 & 8.3 \\
%Blue core & 1.4 & 3.3 & & \\
%Red core  & 1.5 & 3.3 & & \\
%Red wing  & 3.2 & 3.7 & 3.5 & 4.3 \\
Blue wing & 7.2 & 6.0 & 7.5 & 8.3 \\
Blue core & 5.0 & 4.3 & $\cdots$ & $\cdots$ \\
Red core  & 4.9 & 4.3 & $\cdots$ & $\cdots$ \\
Red wing  & 3.6 & 4.4 & 3.5 & 4.3 \\
\hline
\end{tabular}
\label{CCF_results}
\end{table}

\clearpage

\clearpage

%\begin{figure}[p]
%\epsfxsize=\hsize \centerline{\epsfbox{DiskSchematic.epsf}}
%\vspace{.5in}
%\centerline{Figure~1}
%\end{figure}
%
%\begin{figure}[p]
%\epsfxsize=\hsize \centerline{\epsfbox{LineProfiles.epsf}}
%\vspace{.5in}
%\centerline{Figure~2}
%\end{figure}
%
%\begin{figure}[p]
%\epsfxsize=\hsize \centerline{\epsfbox{DiskMap_75_lowres.epsf}}
%\vspace{.5in}
%\centerline{Figure~3}
%\end{figure}
%
%\begin{figure}[p]
%\epsfxsize=\hsize \centerline{\epsfbox{EchoImage_75_lowres.epsf}}
%\vspace{.5in}
%\centerline{Figure~4}
%\end{figure}
%
%\begin{figure}[p]
%\epsfxsize=\hsize \centerline{\epsfbox{ResponseFunctions_75.epsf}}
%\vspace{.5in}
%\centerline{Figure~5}
%\end{figure}
%
%\begin{figure}[p]
%\epsfxsize=\hsize \centerline{\epsfbox{ContinuumLC.epsf}}
%\vspace{.5in}
%\centerline{Figure~6}
%\end{figure}
%
%\begin{figure}[p]
%\epsfxsize=\hsize \centerline{\epsfbox{WingCoreLCs_75.epsf}}
%\vspace{.5in}
%\centerline{Figure~7}
%\end{figure}
%
%\begin{figure}[p]
%\epsfxsize=\hsize \centerline{\epsfbox{CCFs.epsf}}
%\vspace{.5in}
%\centerline{Figure~8}
%\end{figure}

\end{document}